# *An Adaptive GViT for Gas Mixture Identification and Concentration Estimation*

Ding WANG
College of Electronics and Information Engineering
Tongji University
Shanghai, 201804, China
2030764@tongji.edu.cn

Wenwen ZHANG*
School of Electrical and Electronic Engineering
Nanyang Technological University
Singapore, 639798, Singapore
College of Science, Institute of Deep-Sea Advanced Equipment Systems
University of Shanghai for Science and Technology
Shanghai, 200093, China
Wenwen.zhang@ntu.edu.sg

***Abstract*—Estimating the composition and concentration of ambient gases is crucial for industrial gas safety. Even though other researchers have proposed some gas identification and concentration estimation algorithms, these algorithms still suffer from severe flaws, particularly in fulfilling industry demands. One example is that the lengths of data collected in an industrial setting tend to vary. The conventional algorithm, yet, cannot be used to analyze the variant-length data effectively. Trimming the data will preserve only steady-state values, inevitably leading to the loss of vital information. The gas identification and concentration estimation model called GCN-ViT(GViT) is proposed in this paper; we view the sensor data to be a one-way chain that has only been downscaled to retain the majority of the original information. The GViT model can directly utilize sensor arrays' variable-length real-time signal data as input. We validated the above model on a dataset of 12-hour uninterrupted monitoring of two randomly varying gas mixtures, CO-ethylene and methane-ethylene. The accuracy of gas identification can reach 97.61%, $R^2$ of the pure gas concentration estimation is above 99.5% on average, and $R^2$ of the mixed gas concentration estimation is above 95% on average.**

## I. INTRODUCTION

An electronic nose with an artificial intelligence algorithm has been introduced to more and more fields such as biology and medicine [1], food engineering [2], environmental control [3], and so on. One of the most important applications in these areas is the identification of mixed gases. The emission of a gas mixture into the atmosphere also has the potential to be a contributory factor in the occurrence of accidents in the industry.

Gas recognition and concentration estimation of mixed gases have been performed using dictionary learning (DL), Recurrent Neural Network (RNN), and other algorithms. Aixiang He *et al.* proposed a novel DL classification model to enhance the performance of gas identification. [4] Despite having an average precision of over 99 percent, this method is impractical in the industry due to the absence of a uniform data format. A new technique called 2L-ARNN, which uses RNN as its backbone, has already been demonstrated by Xiaofang Pan *et al.* [5]. It is capable of real-time monitoring of gas changes. This model can estimate concentrations while simultaneously classifying gases. However, the majority of the variations in the data they processed only involved concentration, and the output of this model is unknown when the gas composition varies frequently.

While producing good results, the algorithms mentioned above also have fatal flaws. Two aspects of the primary issues can be concluded:

Because changes in gas composition or concentration may occur, the data in a dynamic state cannot be reliably identified if each data point is evaluated independently. The accuracy of gas identification will be reduced when the gas composition changes frequently. Furthermore, some researchers consider the correlation between two consecutive phases of data. However, it only has positive effects if the gas composition does not change. Such occurrences are infrequent in industrial settings. It implies that gas changes at random, and when the algorithm misinterprets the composition of the gas in the dynamic state, it might even have a negative impact on subsequent identification.

We assume that a one-way chain can be used to visualize the data curves from gas sensors. The temporal features of the chain reveal its unidirectionality. When the composition and concentration of the gas are identical, the data gathered earlier has a substantial influence on the data collected afterward. Time can be seen in this light as an attribute of the sensor data. Potentially directional topology structure exists for the temporal properties. The sensor data is therefore regarded as a one-way chain by us.

A graph feature extractor called GCN was introduced by Kipf & Welling.[6] Since it was first proposed, GCN has found widespread application in several disciplines, including network defense[7], neural network fault diagnosis[8], and hyperspectral imaging[9]. In these fields, a topology is a typical form of representation. From our perspective, time chains can be viewed as topologies, allowing GCN to evaluate them.

This article employs a GCN-based method for analyzing a mixture of gases in conjunction with the previously discussed. GCN-ViT (GViT) is employed to determine the gas composition and concentration levels. The significant contributions of this paper are as follows:

1) Existing models cannot analyze sensor data with variant length. This study represents one of the first attempts to fill this critical void by adopting a GCN-based model called GViT. It transcends the limitations of conventional algorithms for adaptively processing data in real time. The data is segmented based on sensor data characteristics rather than artificially defined segmentation. Furthermore, our model is more accurate than existing models at identifying gas composition and estimating concentration. The accuracy of gas identification was improved to 97.61%. The average $R^2$ for estimating the concentration of mixed gases was increased to greater than 95%.

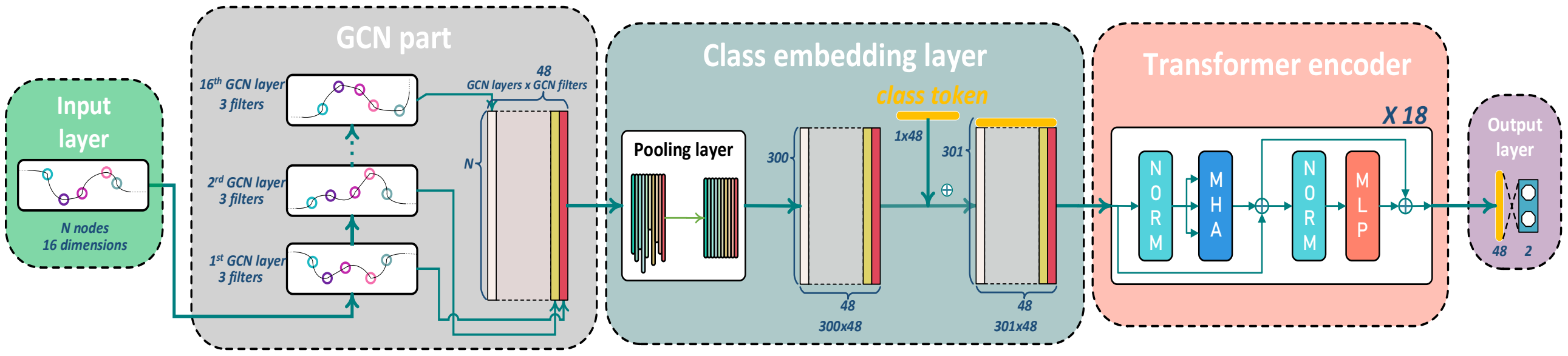

Figure 1 GViT model

2) Due to their lack of generalization, traditional models cannot cope with different mixtures of gases. Changing application conditions frequently necessitates training models with alternative structures. We estimated the concentrations of several gas mixes using the same structural models and achieved good results. Our model is more generalizable and adaptable.

The rest of this article is organized as follows. Section II illustrates the details of data processing and the proposed methodology. Section III presents the experimental verification of the proposed methods and discussion. Section IV is the conclusion.

## II. Data Processing and GViT

### A. Graph of Sensor Array Signal

Some researchers employ time-delay embedding to analyze time series data, such as sensor data. We adopt a different approach, dividing the time series signal $x$ from the sensors based on the label of each time point. The data of each sensor can be viewed as a simplicial complex with one dimension[10]. The overall data can be expressed as follows: $K = \{\{1\},...\{i\},...\{T\},\{1\ ,2\},...\{i\text{-}1\ ,i\},\{i\ ,i{+}1\},...\{T\text{-}1\ ,T\}\}$

Where $\{i\}$ denotes 0-simplices (points), each time point. And $\{i, i{+}1\}$ (edges) denotes 1-simplices, the connection between two data points in a time series. From a topological point of view, $K$ is a straight line carrying the connectivity properties of a time series of length $T$. Based on this, 16 sensor data chains are stacked, giving every point $i$ 16 features.

The equation $\boldsymbol{G} = (\boldsymbol{X}, \boldsymbol{A})$ is frequently used to represent graphs. Where $\boldsymbol{A} \in \mathbb{R}^{N\times N}$ is the adjacency matrix, which depicts the connections between nodes. $A_{ij} = 1$ if there is an edge between $v_i$ and $v_j$; otherwise, $A_{ij} = 0$. $\boldsymbol{X} \in \mathbb{R}^{N\times F}$, where $N$ and $F$ are the number of nodes and features, respectively. In our graph, $\boldsymbol{X}$ represents the matrix of time points, and each time point contains 16 features corresponding to the data of 16 sensors. $\boldsymbol{X}$ and $\boldsymbol{A}$ can be expressed as follows:

$$\boldsymbol{X}_{N\times F} = [\,\boldsymbol{X}_1 ; \boldsymbol{X}_2 ; \ldots ; \boldsymbol{X}_i ; \ldots ; \boldsymbol{X}_N\,] \tag{1}$$

$$\boldsymbol{A}_{N\times F} = \begin{bmatrix} 0 & 1 & & \cdots & & \\ & 0 & 1 & \cdots & & \\ & & 0 & \cdots & & \\ \vdots & \vdots & \vdots & \cdots & \vdots & \vdots \\ & & & \cdots & 0 & 1 \\ & & & \cdots & & 0 \end{bmatrix} \tag{2}$$

Where $X_i = [\, X_i^1,\ X_i^2,\ \cdots,\ X_i^{16}]^T$.

### B. GViT for gas classification and concentration estimation

Vision Transformer employs the encoder module of Transformer to simplify classification by translating image field collections to semantic labels[11]. In light of its success in image classification, we propose GViT for concentration estimation. The significant distinction between GViT and ViT is that GViT uses GCN to extract features. GCN part, class embedding layer, transformer encoder, and final head classifier are the four components that make up GViT. The architecture of GViT is depicted in Fig. 1, where MHA represents multi-head self-attention module, NORM represents the layer normalization module, and MLP represents the multilayer perceptron module.

In the first step, GCN is used to extract the features of the graph G = (X, A). GCN part produces graph matrix, $\boldsymbol{X}_g \in \mathbb{R}^{N\times D}$, where $N$ denotes the number of nodes in the graph, and $D$ denotes the product of the number of GCN layers and filters. Then, a pooling layer is used to combine $N$ nodes into 300 nodes, uniformly forming the shape of each graph. A trainable class token $\boldsymbol{v}_{class}$ is then stacked with graph matrix, where $\boldsymbol{v}_{class} \in \mathbb{R}^{1\times D}$. Class embedding layer produces class matrix, $\boldsymbol{X}_c \in \mathbb{R}^{N'\times D}$, where $N'$ is the number of nodes in the graph following pooling and class token concatenation. The transformer encoder, which consists of $L$ identical blocks with MHA and MLP, receives $\boldsymbol{X}_c$. After extracting class token vclass into the final head classifier, the concentration is then determined, a fully connected layer. Our model has parameters of $D = 48$, $N' = 301$, $L = 18$.

#### 1) Loss

We employ root mean square error (RMSE) as the loss function of GViT model.

$$\text{RMSE} = \sqrt{\frac{1}{2s}\sum_{i=1}^{s}\sum_{g=1}^{2}\left(C_p^{ig} - C_t^{ig}\right)^2} \tag{3}$$

A lower RMSE value indicates a more accurate result, better fitting the actual concentration regression curve,

where $C_p^{ig}$ and $C_t^{ig}$ represent the prediction concentration and target concentration of the *g*-th gas component in the *i*-th sample.

## III. EXPERIMENT RESULTS AND DISCUSSION

### A. Data Descriptions

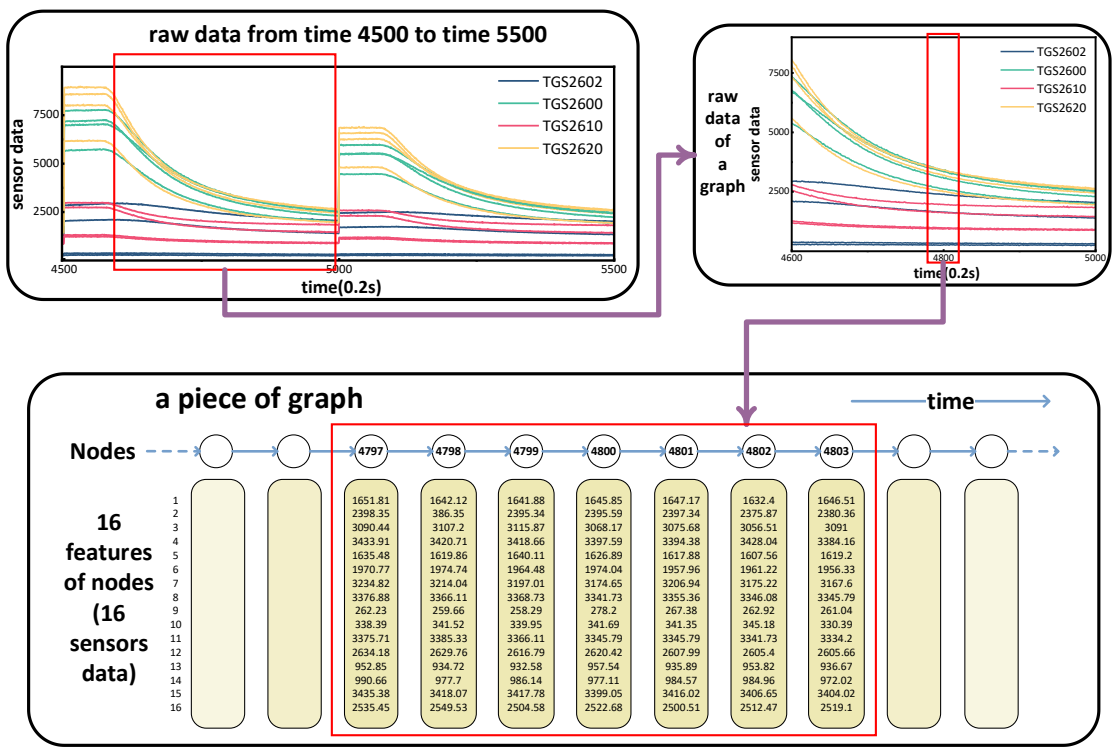

Figure 2 Method of data segmentation

To evaluate the performance of our proposed model for gas identification and concentration estimation, we conducted extensive experiments based on a public dataset from UCI Machine Learning Repository[12]. The dataset contains two gas mixtures: Ethylene and CO in air, Ethylene and Methane in air. Ethylene concentrations vary from 0 to 20 ppm, CO concentrations from 0-533.33 ppm, and methane concentrations from 0-296.67 ppm. The sensor data are constantly recorded at a sampling frequency of 100 Hz for 12 hours. In total, there are 8,387,665 time points. This dataset was downsampled by 5 Hz to eliminate the redundant data, and air phases were omitted. Based on the concentration and composition of the gases, the data was subsequently separated into various shape graphs. As seen in Fig. 2, the image above is a raw representation of an extracted sample transformed into a graph with a topology structure. Each time point can be considered a node in the graph, and the sensing information acquired by each sensor at each instant can be regarded as its attribute. Each node includes sixteen features, as there are sixteen active sensors.

As a consequence of segmentation, the data set has 533 samples. Stratified sampling was used to counteract the effects of sampling imbalance. Before training begins, the test set is divided, and the ratio of the test set is 0.16, making 88 samples total. The details are described in Table I.

TABLE I. DATA DISTRIBUTION

| Group | Gas Composition | The Number Of Dataset | | |
|---|---|---|---|---|
| | | *Total* | *Train-val set* | *Test set* |
| CO + ethylene | CO | 71 | 59 | 12 |
| | ethylene | 111 | 93 | 18 |
| | CO+ethylene | 100 | 84 | 16 |
| methane + ethylene | methane | 76 | 63 | 13 |
| | ethylene | 89 | 74 | 15 |
| | methane+ethylene | 86 | 72 | 14 |

The longest graph has 4351 nodes, equivalent to sensor data that lasts for about 14 minutes; the shortest graph has just five nodes, identical to sensor data that lasts for one second. The 5-fold cross-validation method was used on the training-validation set. The model we tested in each epoch was based on the effectiveness of validation.

We established a base value following information tested in the air to mitigate the impact of the shift of sensors. Additionally, all concentrations were normalized for the concentration estimation task.

$$y_i = \frac{y_{i-original}}{\max(y)} \quad (4)$$

Where $y_{\text{i-original}}$ denotes the *i*-class initial gas concentration, max(*y*) is the maximum concentration of each gas in the dataset.

### B. Results of Gas Classification

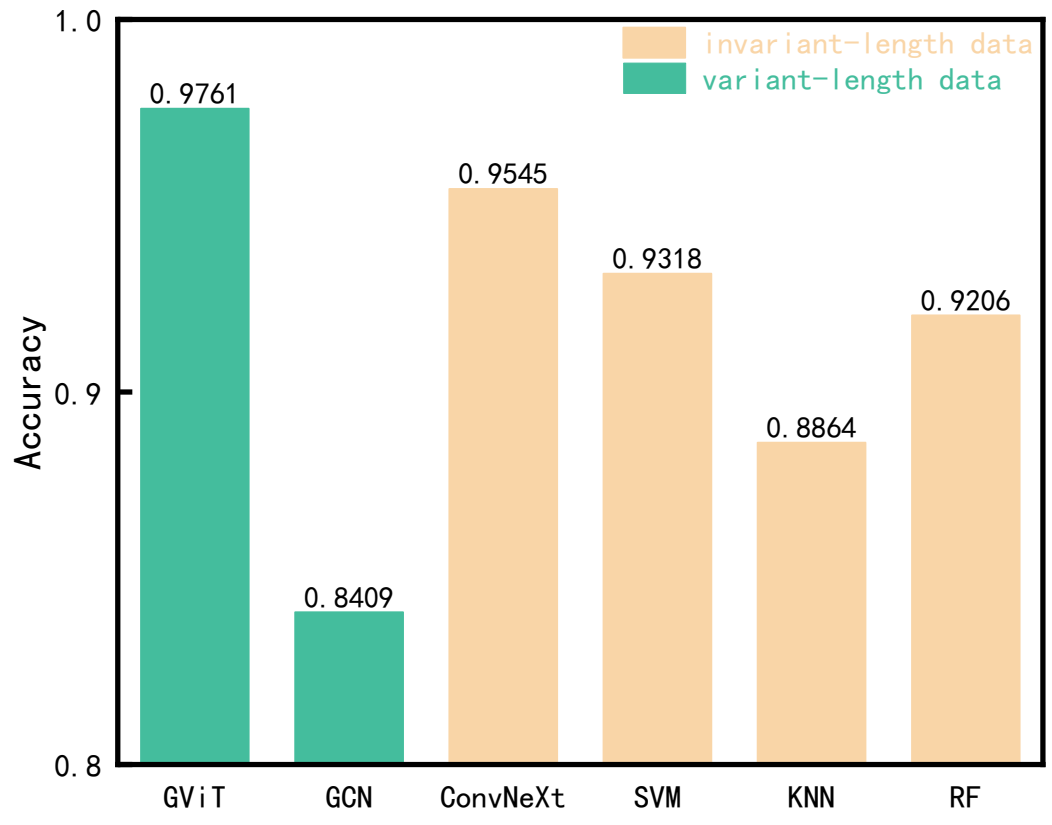

Figure 3 The accuracy of gas identification

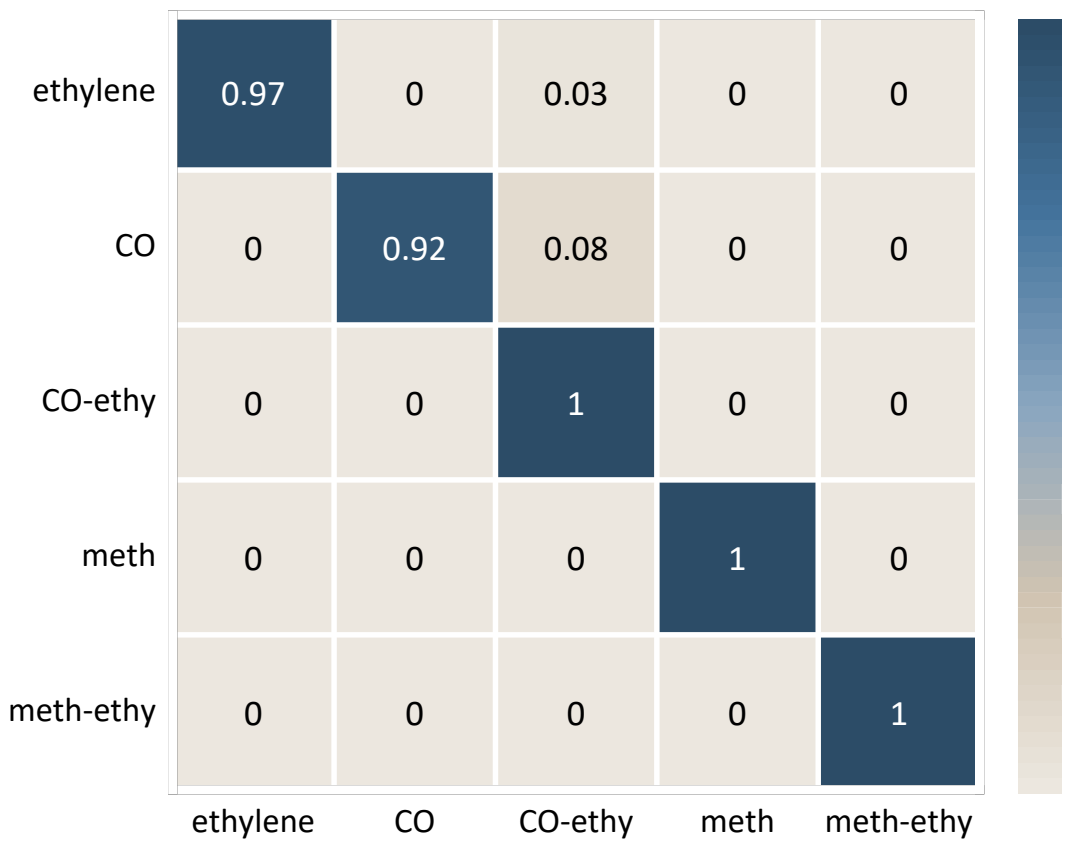

Figure 4 Confusion matrix of GViT

We separately tested two gas mixtures, CO-ethylene and methane-ethylene. The model produces two values representing the concentrations of two different gases inside the group. When the normalized prediction concentration is less than 0.01, we assume that there is no such gas.

To illustrate the performance of GViT, we compare it to various models, including SVM, RF, KNN, GCN and ConvNeXt. [13] Since most of these models cannot accommodate varying data lengths, we only used steady-state data for our evaluations. Since the shortest graph, which we discussed above, only has five nodes, we set all steady-state data to have a length of five nodes. The classification performance of each model is shown in Fig.3. The classification performance of GViT is superior to others, with an accuracy of 97.61%.

Compared to the accuracy of GViT, ConvNeXt scores around two percentage points worse, about 95%. Compared to other models, the identification accuracy of GCN is significantly lower at only 84.09%.

The confusion matrix for gas classification is depicted in Fig. 4. As seen, the data in the confusion matrix of our model are concentrated around the diagonal. The model mistakes only one graph of pure CO and one graph of pure ethylene as its mixture.

### C. *Results of Concentration Estimation*

TABLE II. $R^2$ OF CONCENTRATION ESTIMATION

a) CO + ETHYLENE

| **Model** | **$R^2$** | | | | **RMSE** |
|---|---|---|---|---|---|
| | ***Mixed CO*** | ***Mixed ethylene*** | ***Pure CO*** | ***Pure ethylene*** | |
| GViT | 0.9172 | 0.9561 | 0.9966 | 0.9921 | 0.0213 |
| ViT | -0.3234 | -0.1022 | 0.9519 | 0.8716 | 0.0688 |
| SVM | 0.5112 | -0.8384 | 0.9460 | 0.9630 | |
| RF | 0.0853 | 0.3626 | 0.9654 | 0.9934 | |
| KNN | -0.1735 | -0.3371 | 0.9712 | 0.9355 | |

b) METHANE + ETHYLENE

| **Model** | **$R^2$** | | | | **RMSE** |
|---|---|---|---|---|---|
| | ***Mixed methane*** | ***Mixed ethylene*** | ***Pure methane*** | ***Pure ethylene*** | |
| GViT | 0.9609 | 0.9768 | 0.9994 | 0.9975 | 0.0117 |
| ViT | 0.3517 | 0.2549 | 0.9682 | 0.8906 | 0.0543 |
| SVM | 0.3126 | 0.3689 | 0.9766 | 0.9857 | |
| RF | 0.6587 | 0.1673 | 0.9635 | 0.9860 | |
| KNN | 0.3665 | 0.6326 | 0.9848 | 0.9028 | |

Table II illustrates $R^2$ among algorithms, SVM, KNN, RF and ViT. This table shows that our algorithm has the highest $R^2$ and the lowest RMSE value in both mixed gas groups. $R^2$ of pure gas concentration estimation is much higher than the same gas in mixed gas, as pure gas concentration is much simpler to estimate. In the CO-ethylene group, $R^2$ of mixed gas concentration estimation, CO and ethylene, are 0.9172 and 0.9561, respectively. $R^2$ of pure CO concentration estimation is 0.9966, while $R^2$ of pure ethylene concentration estimation is 0.9921. In another group, $R^2$ of mixed gas concentration estimation are respectively 0.9609 and 0.9768. $R^2$ of pure methane concentration estimation is 0.9994, and $R^2$ of pure ethylene concentration estimation is 0.9975. The concentration estimation $R^2$ of mixed gas in the methane-ethylene group are slightly higher than in the CO-ethylene group. As illustrated in Fig.5, these relationships may partly be explained by the fact that, compared to methane, CO in mixed gas has more concentration levels, with a total of 25 concentration levels between 117ppm and 394ppm, while methane in mixed gas has 11 concentration levels between 74ppm and 181ppm. More concentration levels result in richer variation, which hinders the training of small sample data. Despite this, we still get a good result. We can see in Fig.5 that the 95% Confidence Interval of CO is not much different from the other three gases. We also tested the model ViT. Concentration estimation $R^2$ of ViT can be found in Table II. The results prove that without feature extraction by GCN, ViT cannot identify gas concentration precisely. Its performance is even worse than traditional algorithms.

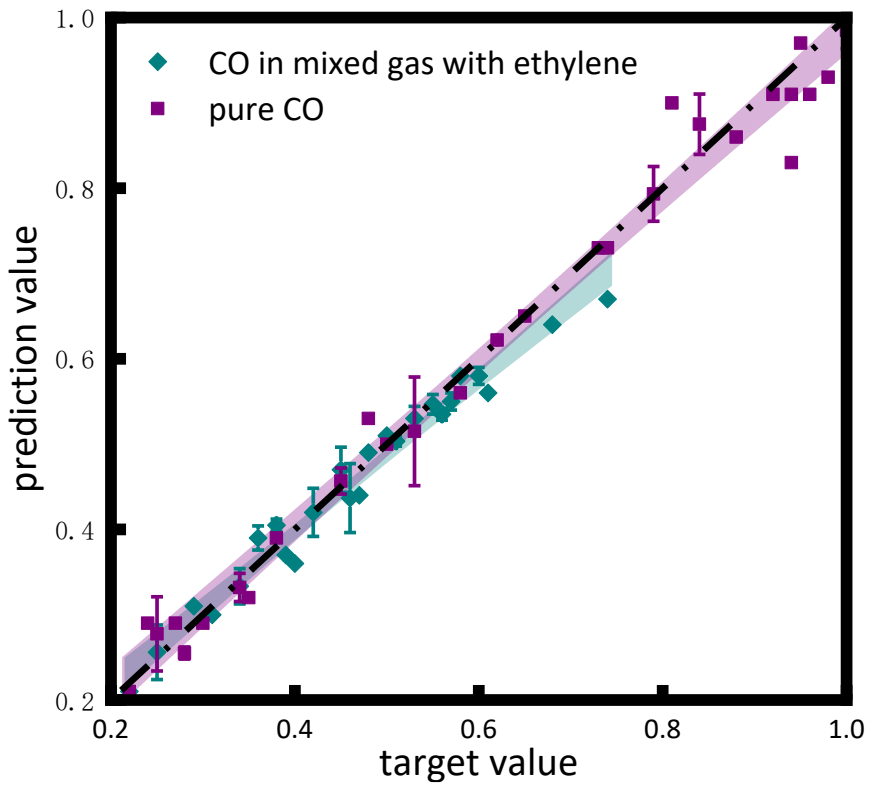

(a)

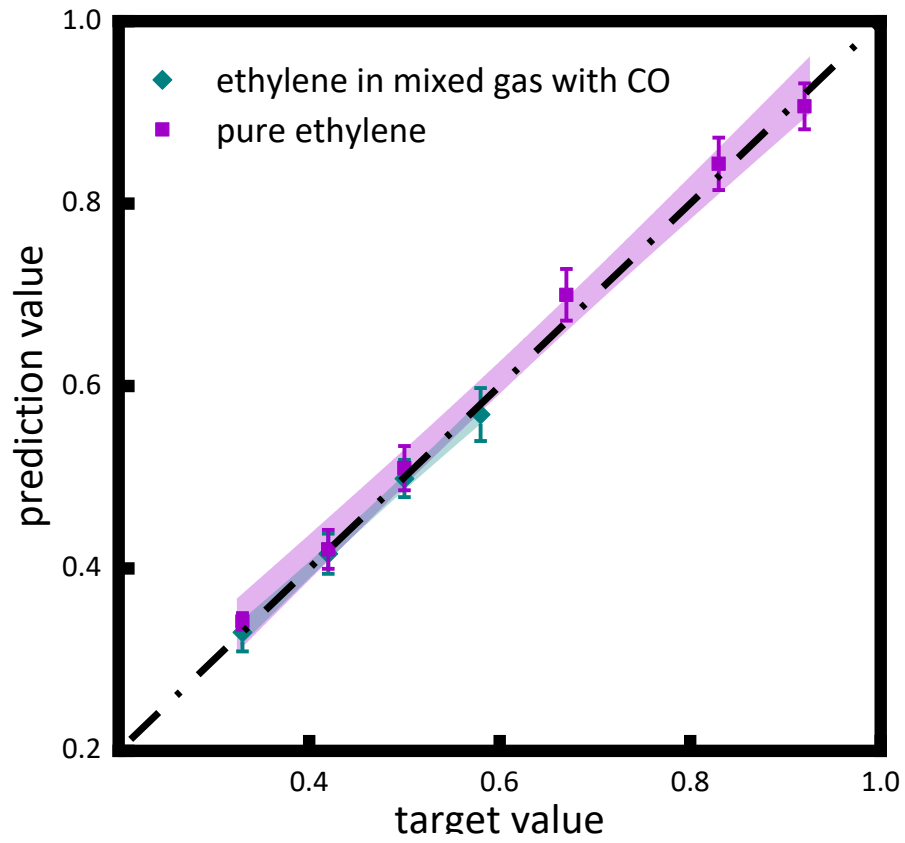

(b)

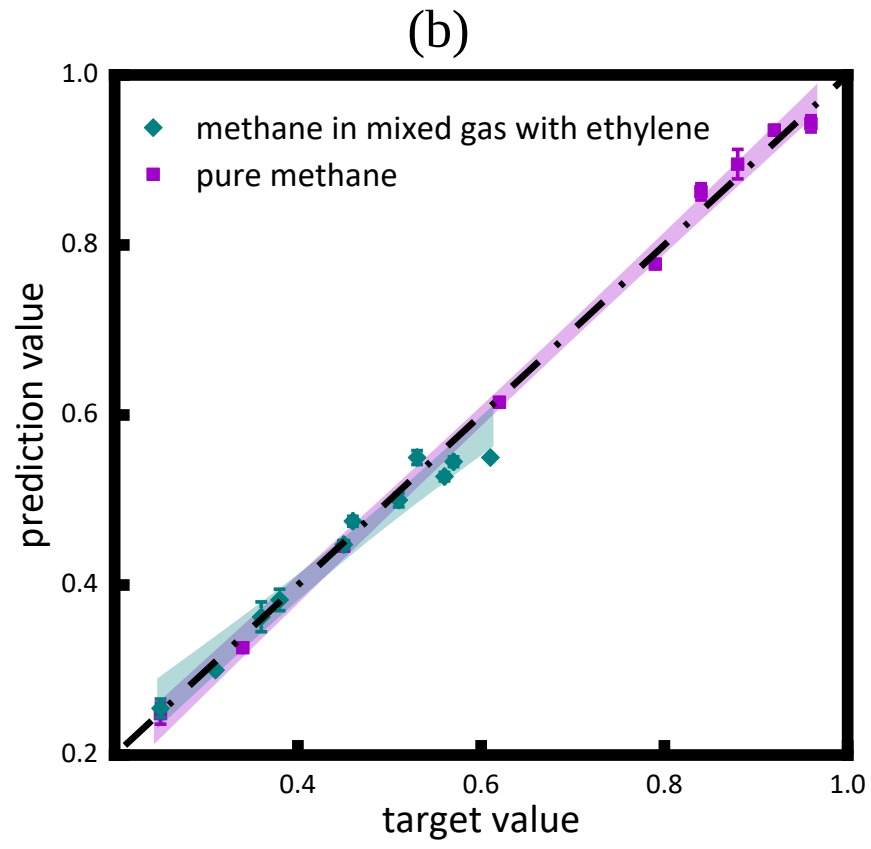

(c)

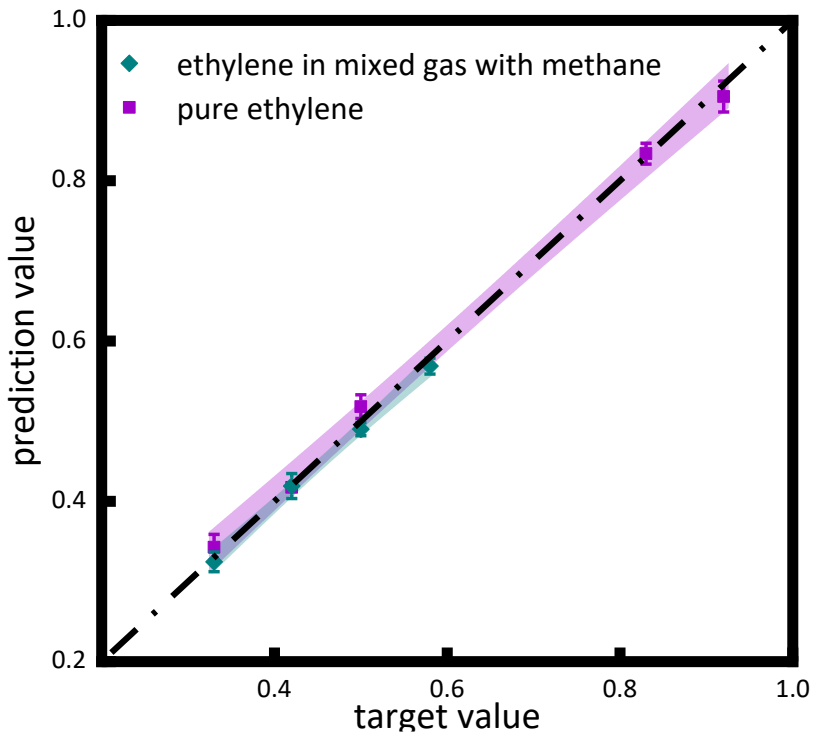

(d)

Figure 5 The error map of concentration estimation. (a) ethylene in group CO-ethylene (b) CO in group CO-ethylene (c) ethylene in group methane-ethylene (d) methane in group methane-ethylene.

## IV. CONCLUSION

This paper addresses the problem that conventional gas identification models and their concentration models cannot directly utilize dynamic response real-time signals with variable length as input. It is challenging to develop a model with good concentration estimation accuracy for all target gases. The gas recognition and concentration estimation model called GViT is proposed in this paper. GViT views the real-time dynamic response signal of variable length obtained from the gas sensor array as a one-way chain serving as input. The proposed model can achieve an accuracy of 97.61% for real-time gas identification. The experimental results show that, compared to traditional concentration estimation models, the GViT model can also achieve much higher concentration estimation accuracy for all target gases. In particular, the gas concentration estimation of each component of the real-time mixed gas is achieved with an average $R^2$ greater than 95%.

## ACKNOWLEDGMENT

This work was supported in part by National Natural Science Foundation of China (No. 6220021889). The authors deeply appreciate the support.

## REFERENCES

[1] B. Day, C.Wilmer "Computational Design of MOF-Based Electronic Noses for Dilute Gas Species Detection: Application to Kidney Disease Detection" ACS Sens., vol. 6, no. 12, pp. 4425-4434, Dec., 2021, doi: 10.1021/acssensors.1c01808.

[2] C. Zhao, J. Shen "Ultra-efficient trimethylamine gas sensor based on Au nanoparticles sensitized WO3 nanosheets for rapid assessment of seafood freshness" Food Chem., vol. 392, Oct., 2022, doi: 10.1016/j.foodchem.2022.133318.

[3] W.Gao, X.Yang "The variation of odor characteristics of wastewater sludge treated by advanced anaerobic digestion (AAD) and the contribution pattern of key odorants" Sci. Total Environ., vol. 840, Sep., 2022, doi: 10.1016/j.scitotenv.2022.156722.

[4] A.He, G.Wei, J.yu "A Novel Dictionary Learning Method for Gas Identification With a Gas Sensor Array" IEEE Trans. Ind. Electron., vol. 64, no. 12, pp. 9709-9715, Dec., 2017, doi: 10.1109/TIE.2017.2748034

[5] X. Pan, Z. Zhang, H. Zhang, Z. Wen, W. Ye, Y. Yang, et al. "A fast and robust mixture gases identification and concentration detection algorithm based on attention mechanism equipped recurrent neural network with double loss function" Sens. Actuators, B, vol. 342, Sep., 2021, doi: 10.1016/j.snb.2021.129982.

[6] T.Kipf, M.Welling "Semi-Supervised Classification with Graph Convolutional Networks", 2016, arXiv:1609.02907.

[7] A. Liu, B. B. Li, T. Li, P. Zhou and R. Wang "AN-GCN: An Anonymous Graph Convolutional Network Against Edge-Perturbing Attacks" IEEE Trans. Neural Netw. Learn. Syst., May, 2022, doi: 10.1109/TNNLS.2022.3172296.

[8] Z. W. Chen, J. M. Xu, T. Peng and C. H. Yang "Graph Convolutional Network-Based Method for Fault Diagnosis Using a Hybrid of Measurement and Prior Knowledge" IEEE Trans. Cybern., Mar., 2021, doi: 10.1109/TCYB.2021.3059002.

[9] J. Y. Yang, H. C. Li, W. S. Hu, L. Pan and Q. Du "Adaptive Cross-Attention-Driven Spatial-Spectral Graph Convolutional Network for Hyperspectral Image Classification" IEEE Geosci. Remote. Sens. Lett., vol. 19, 2022, doi: 10.1109/LGRS.2021.3131615.

[10] Zeng, S., Graf, F., Hofer, C., and Kwitt, R., "Topological Attention for Time Series Forecasting",arXiv e-prints, 2021.

[11] A.Dosovitskiy, "An Image is Worth 16x16 Words: Transformers for Image Recognition at Scale", 2020, arXiv:2010.11929

[12] J. Fonollosa, S. Sheik, R. Huerta and S. Marco "Reservoir computing compensates slow response of chemosensor arrays exposed to fast varying gas concentrations in continuous monitoring" Sens. Actuators, B, vol. 215, pp. 618-629, Aug., 2015, doi: 10.1016/j.snb.2015.03.028.

[13] Liu, Z., Mao, H., Wu, C.-Y., Feichtenhofer, C., Darrell, T., and Xie, S., “A ConvNet for the 2020s”, 2022, arXiv:2201.03545.

.